\documentclass[prb,twocolumn,superscriptaddress]{revtex4-1}
\usepackage{graphicx}
\usepackage{dcolumn}
\usepackage{bm}
\usepackage{color}
\usepackage{amsmath, amsthm, amssymb,stmaryrd}
\usepackage{polski}
\usepackage[polish,french,english]{babel}
\newcommand{\rto}[1]{$R_2$Ti$_2$O$_7${#1}}
\newcommand{\tto}[1]{Tb$_2$Ti$_2$O$_7${#1}}
\newcommand{\hto}[1]{Ho$_2$Ti$_2$O$_7${#1}}
\newcommand{\dto}[1]{Dy$_2$Ti$_2$O$_7${#1}}

\newcommand{\ybto}[1]{Yb$_2$Ti$_2$O$_7${#1}}

\usepackage{gensymb}
\usepackage{amssymb}
\usepackage{epstopdf}
\newcommand{\ra}[1]{\renewcommand{\arraystretch}{#1}}

\begin{document}

\title{First principles calculation and experimental investigation of lattice dynamics in the rare earth pyrochlores $R_2$Ti$_2$O$_7$ ($R =$ Tb, Dy, Ho)}
\author{M Ruminy}
\email{martin.ruminy@gmx.de}
\affiliation{Laboratory for Neutron Scattering and Imaging, Paul Scherrer Institut, 5232 Villigen PSI, Switzerland}
\author{M N\'u\~nez Valdez}
\altaffiliation{Now at: Moscow Institute of Physics and Technology, Dolgoprudny, Moscow Region, Russia}
\affiliation{Materials Theory, ETH Zurich, Wolfgang-Pauli-Str. 27, 8093 Zurich, Switzerland}
\author{B Wehinger}
\affiliation{Laboratory for Neutron Scattering and Imaging, Paul Scherrer Institut, 5232 Villigen PSI, Switzerland}
\affiliation{Department of Quantum Matter Physics, University of Geneva, 24, Quai Ernest-Ansermet, 1211, Geneva 4, Switzerland}
\author{A Bosak}
\affiliation{ESRF - The European Synchrotron, CS40220, 38043 Grenoble Cedex 9, France}
\author{D T Adroja}
\affiliation{ISIS Facility, Rutherford Appleton Laboratory, Chilton, Didcot, Oxon OX11 0QX, United Kingdom}
\affiliation{Highly Correlated Electron Group, Physics Department, University of Johannesburg, P.O. Box 524, Auckland Park 2006, South Africa}
\author{U Stuhr}
\affiliation{Laboratory for Neutron Scattering and Imaging, Paul Scherrer Institut, 5232 Villigen PSI, Switzerland}
\author{K Iida}
\affiliation{Comprehensive Research Organization for Science and Society (CROSS), Tokai, Ibaraki 319-1106, Japan}
\author{K Kamazawa}
\affiliation{Comprehensive Research Organization for Science and Society (CROSS), Tokai, Ibaraki 319-1106, Japan}
\author{E Pomjakushina}
\affiliation{Laboratory for Scientific Developments \& Novel Materials, Paul Scherrer Institut, 5232 Villigen PSI, Switzerland}
\author{D. Prabakharan}
\affiliation{Department of Physics, University of Oxford, Clarendon Laboratory, Parks Road, Oxford, OX1 3PU, UK} 
\author{M K Haas}
\altaffiliation{Now at Air Products and Chemicals Inc., Allentown PA 18195 USA}
\affiliation{Department of Chemistry, Princeton University,  Princeton NJ 08540, USA}
\author{L Bovo}
\affiliation{London Centre for Nanotechnology, University College London, 17-19 Gordon Street, London, WC1H 0AH, UK}
\author{D Sheptyakov}
\affiliation{Laboratory for Neutron Scattering and Imaging, Paul Scherrer Institut, 5232 Villigen PSI, Switzerland}
\author{A Cervellino}
\affiliation{Swiss Light Source, Paul Scherrer Institut, 5232 Villigen PSI, Switzerland}
\author{R J Cava}
\affiliation{Department of Chemistry, Princeton University,  Princeton NJ 08540, USA}
\author{M Kenzelmann}
\affiliation{Laboratory for Scientific Developments \& Novel Materials, Paul Scherrer Institut, 5232 Villigen PSI, Switzerland}
\author{N A Spaldin}
\affiliation{Materials Theory, ETH Zurich, Wolfgang-Pauli-Str. 27, 8093 Zurich, Switzerland}
\author{T Fennell}
\email{tom.fennell@psi.ch}
\affiliation{Laboratory for Neutron Scattering and Imaging, Paul Scherrer Institut, 5232 Villigen PSI, Switzerland}

\date{\today}

\begin{abstract}

We present a model of the lattice dynamics of the rare earth titanate pyrochlores $R_2$Ti$_2$O$_7$ ($R$ = Tb, Dy, Ho), which are important materials in the study of frustrated magnetism.  The phonon modes are obtained by density functional calculations, and these predictions are verified by comparison with scattering experiments.  Single crystal inelastic neutron scattering is used to measure acoustic phonons along high symmetry directions for $R$ = Tb, Ho; single crystal inelastic x-ray scattering is used to measure numerous optical modes throughout the Brillouin zone for $R$ = Ho; and powder inelastic neutron scattering is used to estimate the phonon density of states for $R$ = Tb, Dy, Ho.  Good agreement between the calculations and all measurements is obtained, allowing confident assignment of the energies and symmetries of the phonons in these materials under ambient conditions.  The knowledge of the phonon spectrum is important for understanding spin-lattice interactions, and can be expected to be transferred readily to other members of the series to guide the search for unconventional magnetic excitations.

\end{abstract}

\maketitle

\section{\label{sec:Introduction}Introduction}

The title compounds $R_2$Ti$_2$O$_7$ ($R =$ Tb, Dy, Ho) are three of the most well studied realizations of geometrical frustration~\cite{Gardner:2010fu}.  They support a long running experimental and theoretical quest for understanding of an apparently highly frustrated state when none is expected ($R =$ Tb), and the canonical examples of the spin ice state~\cite{Bramwell:2001tpa} with attendant emergent magnetic monopole excitations ($R =$ Dy, Ho)~\cite{Castelnovo:2008hb}.  Little is known about their lattice dynamics, though these are of potential importance for different reasons.  In \tto{}, the formation of the low temperature state is accompanied by numerous anomalies in elastic properties~\cite{MAMSUROVA:1988wg,Ruff:2007hf,Nakanishi:2011bz}, and, most recently, the hybridization of magnetic and lattice excitations has been advanced as a source of the fluctuations required to melt long-range magnetic order~\cite{Fennell:2014gf}.  In the spin ices, the monopole excitations must hop by reversing large single-ion magnetic moments~\cite{Tomasello:2015vg}, and mechanisms involving interaction between crystal field states and phonons could well play a role~\cite{Finn:2002fs}.  Thus far, the interaction of a crystal field level with a transverse acoustic phonon has been documented in \tto{}~\cite{Fennell:2014gf}, but since the relatively low symmetry of the rare earth site ($D_{3d}$) splits the ground state terms of the Tb$^{3+}$, Dy$^{3+}$ and Ho$^{3+}$ into several levels that are spread over a similar total bandwidth to that typical of acoustic and optical phonons, further interactions may well be possible.   A prerequisite for the understanding of such processes is to know the energy and symmetry of phonon modes which may be involved.

Investigations of pyrochlore-structured materials using electronic structure calculations have mainly been related to their potential applications as host materials to deposit actinides~\cite{Ewing:2004ga}, and as thermal barrier coating materials due to their surprisingly low thermal conductivity at high temperatures~\cite{Schelling:2004fa, Clarke:2005kg}.  For the former, first-principles calculations were used in the study of defect formation in the pyrochlore structure, while for the latter thermodynamic properties were simulated using both the molecular dynamics method~\cite{Schelling:2004fa} and \textit{ab initio} calculations~\cite{Liu:2010he,Guo:2014fy,Yang:2016ih}. For both applications, it was found that pyrochlore zirconates are generally favorable over titanates, hence more theoretical investigations on the lattice dynamics of zirconates have been carried out. In the pyrochlore titanates, available calculations of the phonon spectrum are limited to the $\Gamma$-point, where the energies and symmetries of phonons have previously been predicted~\cite{Kumar:2012,Chernyshev:2015}. Experimentally, the lattice dynamics of both titanate and zirconate pyrochlores were measured using the $\vec{Q}=0$ sensitive Raman scattering \cite{Lummen:2008fh,Maczka:2009ep, SCHEETZ:1979fe} and infrared absorption techniques \cite{Subramanian:1983}, but neither could confirm the existence of the low-lying optical phonon modes that were thought to be responsible for the low thermal conductivity of pyrochlore materials at elevated temperatures \cite{Lan:2015ba}. 

Here we use density functional calculations to predict the entire phonon band structure of \tto{} and \hto{}.  Spectroscopic techniques with finite momentum transfers such as inelastic neutron scattering (INS) and inelastic x-ray scattering (IXS) are needed to determine the phonon dispersion relations across the Brillouin zone in a single crystal, or to collect neutron-weighted powder averages of the phonon density of states (phonon DOS), and these techniques are then used to validate our calculations and symmetry assignments.  We find good agreement with our model throughout the Brillouin zone.  By comparing the experimentally determined phonon DOS in \tto{}, \dto{}, and \hto{}, we find that the phonon frequencies evolve only gradually across the series, and so our dispersions and symmetries may be taken as a good guide for understanding excitations in other nearby compounds such as \ybto{}~\cite{Pan:2014jz}.  After introducing our computational (\ref{sec:Computational_methods}) and experimental (\ref{sec:Experimental_methods}) methods, the paper presents detailed results concerning structural relaxation (\ref{sec:relax}) and the calculation of the phonon band structure (\ref{calc_phonons}); the verification of these predictions by inelastic neutron scattering (\ref{sec:ins_cryst} and \ref{sec:ins_pow}), and inelastic x-ray scattering (\ref{sec:ixs}); followed by discussion (\ref{sec:Discussion}) and conclusions (\ref{sec:Conclusion}).  Sample parameters and calculated lattice heat capacities of \tto{} and \hto{} can be found in the appendix.  Readers interested only in general features of the phonon band structure of rare earth titanates will find an overview of the dispersion relations and partial phonon DOS of \hto{} in Fig.~\ref{fig:hto_phonons} and Fig.~\ref{fig:theory_pphdos} respectively, and a tabulation of energies and symmetries of zone center phonons in Table~\ref{tab:phonon_symmetries}.

\section{\label{sec:Computational_methods}Computational methods}

We have applied density functional theory within the Perdew-Burke-Ernzerhof (PBE) \cite{Perdew:1996} parametrized generalized gradient approximation (GGA) optimized for solids (PBEsol) \cite{Perdew:2008} using the plane-wave basis projector augmented wave (PAW) \cite{Blochl:1994} method as implemented in the VASP code \cite{Kresse:1996-1,Kresse:1996-2,Kresse:1999}. The energy cutoff of the plane-wave basis was checked for convergence of the structural parameters and subsequently fixed to 550\,eV. 

The electronic potentials of the ions were approximated by PAW GGA pseudo-potentials using the electronic valence contribution $6s^25p^65d^1$ for the rare earths, $3p^63d^34s^1$ for titanium and $2s^22p^4$ for oxygen.  The $f$-electrons of the rare earth ions were frozen into the core states, an approach which was used previously \cite{Kumar:2012} and proven not to affect the results of phonon calculations~\cite{Chernyshev:2015}.  The primitive reduced unit cell containing 22 atoms (see Fig.~\ref{fig:primitive_cell}) was sampled by a $4\times4\times4$ $k$-grid generated from the Monkhorst-Pack scheme \cite{Monkhorst:1976}. The total energy was minimized until the differences in the total forces were smaller than $10^{-4}$\,eV/\AA. The atom positions and volume of the reduced unit cell were relaxed at both ambient and applied external pressures to obtain the equilibrium structures.

The phonon calculations for the two rare earth titanates \tto{} and \hto{}  were carried out using the finite displacement method as implemented in the Phonopy code~\cite{Togo:2008,Togo:2015jm}. Distorted atomic configurations in a $2\times2\times2$ supercell containing 176 atoms were generated and the induced forces were calculated by using VASP, with the same precision as employed for the structural relaxation.  The atomic displacement amplitude of 0.01\,\AA{} was verified to give forces that depend linearly on the displacements.  The static dielectric tensor and Born effective charges were calculated using density functional perturbation theory (DFPT) as implemented in VASP. Using Phonopy, non-analytical term corrections were applied to the dynamical matrix at $\vec{Q}\rightarrow0$ and interpolated to general $\vec{Q}$ according to the interpolation scheme by Wang {\it et al.}~\cite{Wang:2010ks}. The total and partial phonon densities of states were evaluated on a $16\times16\times16$ $\Gamma$-centered mesh (whose size was tested for convergence) using the Parlinski-Li-Kawazoe Fourier interpolation scheme~\cite{Parlinski:1997kr} and smeared with a Gaussian of width $\sigma=1.1$~meV.  The lattice heat at constant volume was calculated from the total energy of the phonon bath in the harmonic approximation using a sampling mesh of size $71\times71\times71$, which yielded a convergence of better than $4\times10^{-2}$ at the lowest temperatures.

\section{\label{sec:Experimental_methods}Experimental methods}

\subsection{Neutron scattering (powders)}

We investigated the phonon density of states using inelastic neutron scattering experiments on powders.  The samples were prepared from stochiometric ratios of the oxides Ho$_2$O$_3$, Dy$_2$O$_3$ or Tb$_4$O$_7$, and TiO$_2$ in a solid state reaction.   The oxides, with 99.99\% purity, were annealed at 850 {\degree}C for 10 hours, then mixed and ground, and heated at 950-1300 {\degree}C for 140 hours with several intermediate grindings.  The structures were verified by combined neutron and x-ray diffraction experiments, which were carried out on HRPT~\cite{Fischer:2000to} at SINQ, PSI and the Materials Science Beamline (MSB)~\cite{Willmott:2013gv} at the SLS, PSI.    Rietveld refinement of the structures (results tabulated in Appendix~\ref{appendix:characterization}) as implemented in the Fullprof~\cite{fullprof} software proved all samples to be of high quality and single phase. 

Inelastic neutron time-of-flight measurements on the powder samples of \hto{} and \tto{} were performed on the MERLIN spectrometer at ISIS \cite{Bewley:2006}. The samples (each of mass $\approx 10$ g) were packed in envelopes of aluminum foil which were curled up to form an annular cylinder with diameter and height of 40\,mm. Subsequently, the samples were sealed into aluminum cans containing helium exchange gas, and cooled by a closed-cycle refrigerator on the instrument.  Different settings with incoming neutron energies of $E_i=60$ and $150$\,meV, and corresponding chopper frequencies of $f=400$, and $600$\,Hz were chosen to record data at $T=5, 50,$ and $200$\,K for $400$\,{$\mu$}Amp hrs ($\approx 2$ hours at ISIS full power) each.  The instrumental background was expected to be negligible and hence not measured. The raw data were corrected for detector efficiency by normalizing the intensities using a standard vanadium sample.

\dto{} was investigated using the 4SEASONS spectrometer at J-PARC~\cite{Kajimoto:2011ja}. The sample (mass $\approx 5$ g) was packed in an aluminum foil envelop which was wrapped into a cylinder of 30 mm diameter and 50 mm height, and then sealed in an aluminum can with helium exchange gas.  The thickness of the sample was carefully controlled so as not to exceed 0.5 mm, to optimize the inelastic signal despite the large absorption cross section of natural dysprosium.  4SEASONS was operated in repetition rate multiplication mode~\cite{Nakamura:2009iw}.  Using a Fermi chopper frequency of 250 Hz, the phases of the other choppers were configured so that for a single source pulse, spectra were recorded  for $E_i=153.2, 55.4, 28.3, 17.1$ meV.  Measurements were taken at $T = 200$ K, for 8 hours. The detector pixel efficiency was calibrated using a vanadium standard sample. The instrumental background was measured to subtract the significant contribution from scattering due to phonons of the aluminum sample can from the raw data~\cite{Inamura:2013fe}. 

For the experimental determination of the phonon DOS ($g^{\rm NW}(E)$) in the incoherent approximation, we used data collected on MERLIN ($R=$Tb, Ho) with incoming neutron energies $E_i=60$ and 150\,meV and on 4SEASONS ($R=$Dy) with $E_i=153.2$\,meV.  The low energy (LE) setting was chosen to exploit the better instrumental resolution at lower incoming energies and resulting energy transfers, while the high energy (HE) setting covers the entire spectrum of incoherent one-phonon scattering.  The scattered neutron intensity is integrated over scattering angles $\phi$ ranging from 60 to 80 degrees (HE), or 85 to 135 degrees (LE), of which the latter was only accessible on MERLIN. By integration over the scattering angle we evaluate the neutron-weighted phonon density of states from the same $|\vec{Q}|$-range measured on the two instruments (which do not have identical detector coverage). Magnetic contamination from strong crystal field excitations is excluded by carefully limiting the integration to sufficiently large $\phi$ angles. For each setting, the (incoherent) elastic line was removed and replaced by a Debye extrapolation below 12 meV in the HE setting and 4.5 meV in the LE setting~\cite{Kresch:2008}. Multiphonon and multiple scattering were removed from the signal up to fourth order using the iterative scheme of Sears {\it et al.}~\cite{Sears:1995}, as extended by Kresch {\it et al.} \cite{Kresch:2007kr}. The scaled multiple scattering contribution is found to be close to parity with the multiphonon sum. From the resulting one-phonon scattering profile, $g^{\rm NW}(E)$  is obtained by correcting the thermal phonon occupation following Bose statistics. Eventually, scaled fractions obtained from the different $E_i$ settings were concatenated at 36\,meV, at which energy the phonon DOS peaks. 

\subsection{Neutron spectroscopy (single crystals)}

Neutron measurements of the acoustic phonon dispersion relations in \hto{} and \tto{} single crystals were performed on the thermal triple-axis neutron spectrometer EIGER at the Swiss neutron spallation source SINQ.  The \hto{} sample used was a large single crystal of mass $\approx3$ g, which was grown from a lead fluoride flux~\cite{Garton:1968dz}.  It was originally used to characterize the spin ice state in \hto{}~\cite{Harris:1997uy,Bramwell:2001hm}.  The \tto{} sample was a large single crystal of mass $\approx7$ g, grown in a floating zone furnace.  It has been previously used to investigate diffuse and inelastic neutron scattering~\cite{Fennell:2012ci,Fennell:2014gf}, and characterized extensively~\cite{Fennell:2014gf,mem_all_samps}.  The crystals were both aligned with the $[1\bar{1}0]$ direction vertical, to give an $(h,h,l)$ scattering plane. This configuration allows the measurement of longitudinal and in-plane transverse acoustic phonons along the cubic high symmetry directions. Individual phonon branches were measured in Brillouin zones chosen to satisfy the selection rules for phonon scattering. The final neutron wave vector was usually fixed at $k_f=2.662$\,\AA$^{-1}$ but needed to be increased to $k_f=3.4$\,\AA$^{-1}$ to access phonon excitations in Brillouin zones with large momentum transfers. For $k_f=2.662$\,\AA$^{-1}$ a pyrolytic graphite filter was installed in the scattered beam to eliminate contamination by higher order scattering. For higher final neutron energies, the filter was removed and possible higher-order scattering, dominantly from optical phonons, was considered during the analysis. Phonon excitations were measured at $T=200$\,K with constant energy scans for steep parts of the dispersion branches, and elsewhere with constant $\vec{Q}$-scans.

\subsection{X-ray spectroscopy}

We employed inelastic x-ray scattering (IXS) to access optical phonon branches in \hto{}.  A rectangular rod was cut from a piece of the same boule which supplied the sample used in Ref.~[\onlinecite{Fennell:2009ig}], a floating-zone grown and oxygen annealed single crystal.  The rod was aligned such that the $[1\bar{1}0]$ direction was parallel to the long axis and was polished down to 70\,{$\mu$}m thickness and 600\,$\mu$m length.  Samples used for such measurements may be etched with hydrofluoric acid, but this was found to be unnecessary for \hto{}.

The crystal was mounted on the ID28 beamline at the ESRF, Grenoble. The spectrometer was operated with an incoming photon energy of 17.794\,keV selected by the Si$(9,9,9)$ reflection of the backscattering monochromator. The needle-like sample, which was found to be aligned within two degrees in the $(h,h,l)$-plane, was mounted in a Joule-Thompson cryostat. In this configuration all nine detector positions correspond to $\vec{Q}$-points in the scattering plane, such that we could efficiently measure phonons at nine $\vec{Q}$-points in a single energy scan.

\section{\label{sec:Results}Results}

\subsection{\label{sec:relax}Structural relaxation}

\begin{table}
\caption{The crystallographic positions of the four independent atoms of $R_2$Ti$_2$O$_6$O$'$ in conventional cubic cell with space group $Fd\bar{3}m$ and origin at the Ti site. The $x$ parameter of the O(48$f$) ions is approximately 0.33 for $R=$Tb, Dy, Ho. }
\centering
\ra{1.3}
\begin{tabular}{c c c c c c l}
\hline\hline
Atom  & Fractional   & Wyckoff site & Point group \\
      & coordinates  &              & \\
\hline
R $^{\phantom{`}}$ &  $\frac{1}{2},\frac{1}{2},\frac{1}{2}$ & $16d$ & D$_{3d}$ \\ 
Ti$^{\phantom{`}}$ & $0,0,0$                                & $16c$ & D$_{3d}$ \\
O$^{\phantom{`}}$  & $x,\frac{1}{8},\frac{1}{8}$            & $48f$ & C$_{2v}$ \\
O$^{'}$            & $\frac{3}{8},\frac{3}{8},\frac{3}{8}$  &  $8b$ & T$_{d}$ \\
\hline\hline
\end{tabular}
\label{tab:crystallography}
\end{table}

\begin{figure}
  \centering
   \includegraphics[trim=220 330 200 330,clip=true, scale=0.95]{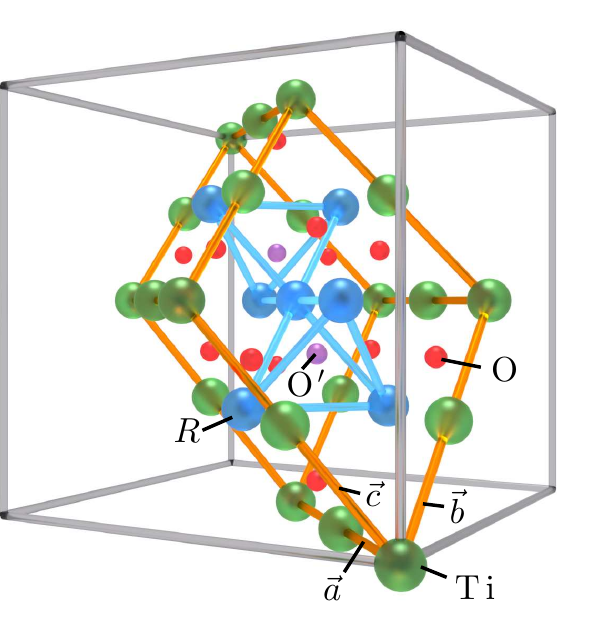}%
  \caption{The primitive cell of the pyrochlore structure, as related to the conventional cubic unit cell.  The primitive cell contains 22 ions: 4 $R^{3+}$ (blue), 4 Ti$^{4+}$ (green), 12 O$^{2-}$ (red) and 2O$'^{2-}$ (violet).  The axes of the conventional cell are shown by the grey box, and the primitive cell by the orange box.  The basis vectors of the primitive cell (in the conventional cell) are $\vec{a}=(1/2,1/2,0)$, $\vec{b}=(1/2,0,1/2)$ and $\vec{c}=(0,1/2,1/2)$.  (The size of the ions is arbitrary.)}
   \label{fig:primitive_cell}
\end{figure}

The rare earth titanate pyrochlores $R^{3+}_2$Ti$^{4+}_2$O$^{2-}_6$O$'^{2-}$ (here $R$ = Tb, Ho) crystallize in the ideal pyrochlore structure with space group $Fd\bar{3}m$ \cite{Gardner:2010fu}, with atom positions as listed in Table \ref{tab:crystallography}.  The primitive unit cell used in the calculations is derived from the conventional unit cell as depicted in Fig.~\ref{fig:primitive_cell}, and contains two formula units with 22 atoms in total.  The pyrochlore structure is controlled by only two parameters which need to be optimized during the self-consistent structural relaxation: the lattice constant $a$, and the $x$-coordinate of the oxygen atoms on the $48f$ site. 

The optimization of the lattice was performed under three different conditions, the results of which are summarized in Table \ref{tab:equi_lat}.  Firstly, using the PBE parametrization of GGA, the lattice volume is overestimated by 0.4\,\% compared to the experimental value, as expected.  Using PBE parameterization tuned for densely packed solid structures (PBEsol), the lattice volume is underestimated by 0.8\,\%; and thirdly, applying an additional isotropic pressure of 50\,kbar, the lattice volume is underestimated by 1.6\,\%, which is a typical value for a structural relaxation performed in the framework of the local density approximation (LDA).  Overall, the calculated values for the lattice constants and $x$-parameters for both \hto{} and \tto{} are in good agreement with both experimental values and DFT calculations using comparable parameterizations \cite{Kumar:2012, Chernyshev:2015}.

\begin{table}
\caption{Comparison of the structural equilibrium parameters obtained in this work with calculations reported in the literature using a variety of exchange-correlation functionals. The experimental values for the lattice parameters are those obtained from powder x-ray diffraction at room temperature (as reported in the Appendix), and corrected for thermal expansion of $\approx2\times10^{-3}$ (as measured for \tto{}
 between 45 and 300 K~\cite{Han:2004bz}). (The relevant experimental parameters for \dto{} are $a=10.105(1)$ \AA ~and $x=0.3278(2)$.)}
\ra{1.5}
\centering
\begin{tabular}{l r r c r r}\hline\hline
Method & \multicolumn{2}{c}{\tto{}}  & &  \multicolumn{2}{c}{\hto{}}  \\
\cline{2-3} \cline{5-6}
 & \multicolumn{1}{c}{$a(\text{\AA})$} &  \multicolumn{1}{c}{$x$} & & \multicolumn{1}{c}{$a(\text{\AA})$} &  \multicolumn{1}{c}{$x$}\\
 \hline
PBE                & 10.1990 & 0.3295 & & $-$     & $-$   \\
PBEsol             & 10.0794 & 0.3303 & & 10.0233 & 0.3318\\
PBEsol, $50$\,kbar & 10.0011 & 0.3309 & & 9.9462  & 0.3324\\
LDA \cite{Kumar:2012}           & $-$    & $-$   & & 9.9301  & 0.3315 \\
PBE0 \cite{Chernyshev:2015}         & 10.171 & 0.328 & & 10.118  & 0.329\\
B3LYP \cite{Chernyshev:2015}        & 10.278 & 0.327 & & 10.222  & 0.328\\
Experiment  & 10.1331(1) & 0.3271(2)  & & 10.082(1) & 0.3285(2)\\
\hline\hline
\end{tabular}
\label{tab:equi_lat}
\end{table}

\subsection{\label{calc_phonons}Phonons from DFT calculations}

\begin{figure}
  \centering
   \includegraphics[scale=1]{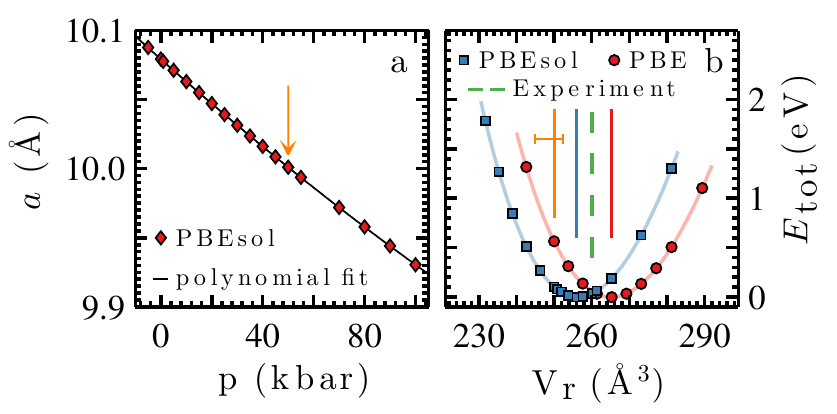}%
  \caption{The application of isotropic pressure resolves artificial phonon instabilities in the PBEsol calculations, shown for \tto{}. Panel a: The pressure dependence of the lattice parameter $a$. The arrow indicates the pressure of 50\,kbar, which was chosen for the calculation. Panel b: Murnaghan equation of state fits  to calculations using PBE and PBEsol parameterizations. The vertical lines indicate the (ambient pressure, corrected for thermal expansion) experimental lattice volumes, and the lattice volumes  of PBE and PBEsol calculations at 50\,kbar. The horizontal orange error bar denotes the expected underestimation of the lattice volume using a LDA exchange-correlation functional, and the orange line indicates that we can reach this lattice volume by applying pressure to move along the equation of state curve. }
   \label{fig:tto_pressure}
\end{figure}

High precision in the structural relaxation with vanishing internal forces on each ion is essential to calculate reliable phonons within the harmonic approximation. For the titanate pyrochlores we identified a further factor that sensitively controls the phonon band structure calculation: the lattice volume.  Although the ionic structure was optimized to reduce the total forces to less than 10$^{-4}$\,eV/\AA{} per atom in all three conditions introduced above - PBE, PBEsol and PBEsol under pressure - the calculated phonon band structure from the theoretical equilibrium values for both PBE and PBEsol parameterizations show unphysical imaginary phonon modes across the entire Brillouin zone.  Experimentally, no sign of structural instabilities of the pyrochlore structure is reported for \rto{} with $R$ = Tb, Dy, Ho. 

Intensive testing showed that these soft modes originate neither from improper matching of the plane wave basis and $k$-point grid, nor insufficient sizes of the supercell or atomic displacements. These artifacts persist when calculating the phonon band structure from relaxed ions in a unit cell fixed by the experimental lattice parameter, and moreover appear to be independent of the A-site ion. Similar artifacts were observed for Y$_2$Ti$_2$O$_7$ when using GGA exchange-correlation functionals (not shown), but vanished when switching to LDA.  Since PAW LDA pseudo-potentials for the rare earth ions were not available, we applied isotropic pressure to squeeze the unit cell towards the theoretical equilibrium lattice parameter of a calculation using LDA exchange-correlation functionals (i.e. underestimating the lattice constant by 1-3\,\%), and this allowed the calculation of  a stable phonon spectrum.  Details of this process are shown in Fig.~\ref{fig:tto_pressure} for \tto{}, and a qualitatively identical behavior was found for \hto{}.  As illustrated in Fig.~\ref{fig:tto_pressure}b, the equation of state allows us to modify the system from the ambient pressure lattice volume obtained from a relaxed structure resulting from a PBEsol or PBE calculation, to obtain relaxed structures with the lattice volume expected for an LDA calculation by application of a small, positive, isotropic pressure.


\begin{figure}
  \centering
   \includegraphics[scale=1]{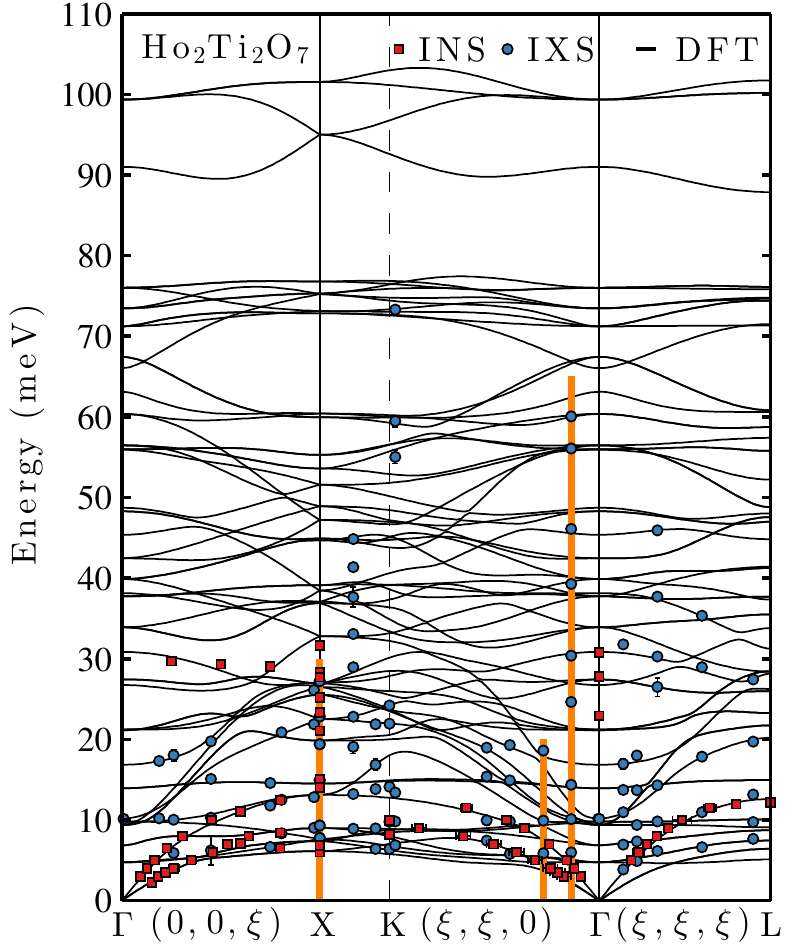}%
  \caption{Phonon dispersion relations of \hto{} calculated using DFT and the finite displacement method. The vibrational spectrum is presented along a path following high symmetry directions of the reciprocal lattice.  The calculation is experimentally verified using inelastic neutron (INS) and x-ray (IXS) scattering. INS and IXS frequencies were obtained from fits to the measured spectra, as described in the text. The INS measurements of the acoustic phonon spectrum are presented in more detail in Fig.~\ref{fig:acoustic_phonons}, and a comparison between simulated and measured IXS intensities along the three broad orange lines is shown in Fig.~\ref{fig:exp_vs_theory_ixs}.}
   \label{fig:hto_phonons}
\end{figure}

Fig.~\ref{fig:hto_phonons} shows the calculated phonon band structure projected on a path along high symmetry directions of the reciprocal lattice for \hto{}. The calculation was performed using the PBEsol parameterization of the GGA exchange-correlation functionals and an applied isotropic pressure of 50\,kbar. As the reduced unit cell contains 22 atoms, the phonon spectrum consists of 66 branches, which are partly degenerate along the high symmetry lines. Along with the calculated phonon dispersions, the extracted phonon dispersion points from INS and IXS measurements are presented (detailed discussion of the comparison with these measurements follows below). The overall agreement between theory and experiment spanning a wide range of momentum and energy transfers is good. The phonon frequencies at the $\Gamma$-point of both \hto{} and \tto{} are summarized in Table \ref{tab:phonon_symmetries}, along with calculations of zone-center phonons available in the literature. In particular the phonon frequencies of \hto{} agree closely with the previous values calculated by Kumar \textit{et al.}~\cite{Kumar:2012} using LDA, which is expected due to the pressure tuning of the unit cell that we employed.
      
\begin{table}
\caption{Symmetries and frequencies (in meV) of zone-center phonons in \rto{}, with $R=$ Ho, Tb, calculated in this work (PBEsol) and in works of Kumar \textit{et al.} (LDA) \cite{Kumar:2012} and Chernyshev \textit{et al.} (PBE) \cite{Chernyshev:2015}. In consequence of the non-analytical term corrections, the vibrational modes with F$_{1u}$ symmetry split into longitudinal optic (LO) and transverse optic (TO) modes with A$_{1u}$ and E$_u$ symmetries, respectively.}
\ra{1.3}
\centering
\begin{tabular}{l c c c c}\hline\hline
Symmetry &  \multicolumn{2}{c}{\hto{}}  &  \multicolumn{2}{c}{\tto{}} \\
\cline{2-3}  \cline{4-5} 
 & PBEsol & LDA \cite{Kumar:2012} & PBEsol & PBE0 \cite{Chernyshev:2015} \\
 \hline
A$_{1g}$ &  63.11 &   63.1    &  62.96 & 65.2   \\ \hline
A$_{2u}$ &  30.85 &    30.7    & 32.06   & 32.0  \\
         &  45.37 &     45.4   & 45.32   & 45.0  \\
         &  48.72 &     49.1   & 48.58   & 56.9  \\ \hline
E$_{g}$  &  42.47 &   42.0     &  42.28   & 40.3  \\ \hline
E$_{u}$  &  9.79  &    9.5    &  10.25   & 9.9 \\
         &  21.23 &     21.4   &  21.98   & 23.7 \\
         &  60.36 &    60.0   &  59.49   & 60.6 \\ \hline
F$_{1g}$ &  33.90 &   34.2    &  34.37   & 33.5 \\
         &  71.24 &    70.5   &  70.00   & 67.1 \\ \hline
F$_{1u}$ (E$_u$, A$_{1u}$) &  $6.86 , 10.20$  & 7.8  &  $ 7.70 , 10.78$ & 12.3 \\
         &  $13.94 , 16.85$ & 13.9 &  $14.62 , 17.25$ & 15.5 \\
         &  $21.13 , 26.75$ & 21.2 & $ 21.85 , 28.02$ & 23.7 \\
         &  $27.44 , 38.13$ & 28.1 &  $28.54 , 38.26$ & 32.7 \\
         &  $48.29 , 55.93$ & 47.5 &  $47.50 , 55.94$ & 46.1 \\
         &  $55.96 , 66.04$ & 55.8 &  $56.13 , 65.24$ & 54.6 \\
         &  $67.44 , 91.01$ & 67.6 & $ 66.56 , 90.15$ & 68.2 \\ \hline
F$_{2u}$ &  4.78     & 4.2  &  4.79  & 4.6 \\
         &  9.38     & 9.9  &  9.26  & 11.9 \\
         &  37.74    & 37.8 &  37.19 & 36.0  \\
         &  76.01    &75.1  &  74.28 & 72.3  \\ \hline
F$_{2g}$ &  39.89    & 39.7 &  40.04 & 38.4  \\  
         &  56.46    & 56.4 &  56.26 & 56.0  \\
         &  73.47    & 73.4 &  73.87 & 72.8  \\ 
         &  99.35    & 99.9 &  98.15 & 98.8  \\        
\hline\hline
\end{tabular}
\label{tab:phonon_symmetries}
\end{table}

\begin{figure}
  \centering
   \includegraphics[scale=1]{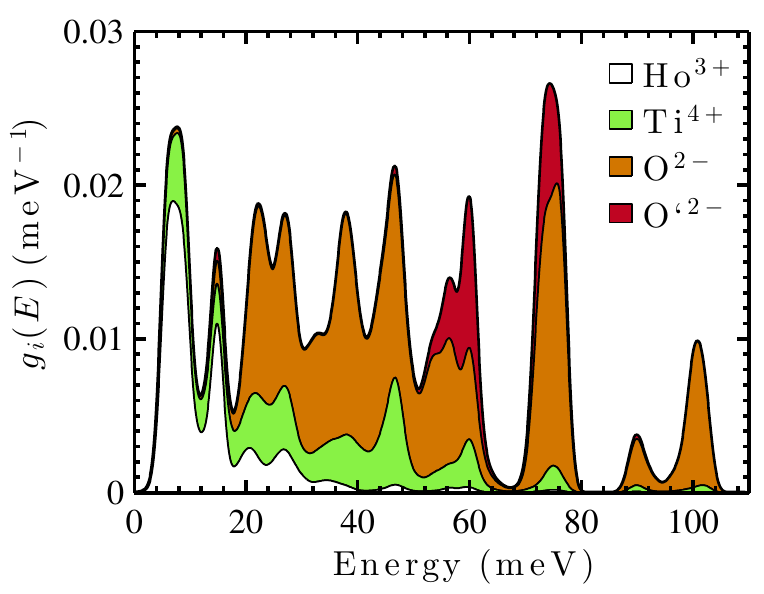}%
  \caption{Normalized partial phonon densities of states $g_i(E)$ of \hto{}, calculated from first-principles. }
   \label{fig:theory_pphdos}
\end{figure}

The calculated phonon DOS of \hto{} is shown in Fig.~\ref{fig:theory_pphdos}, and provides a compact way to visualize three important general aspects of the rare earth titanate phonon spectrum.  Firstly, the phonons spread over the same energy range as the splitting of the crystal field ground state multiplet in rare earth titanate pyrochlores, which forms the basis for magnetoelastic interactions in these materials (some of which have been reported elsewhere~\cite{Fennell:2014gf}. Breaking down the phonon DOS into the partial contributions due to the four independent ions of the pyrochlore structure, we find that the heavy rare earth ions contribute to lower frequency modes, in contrast to the lighter oxygen ions, which dominate the phonon vibrations above 20\,meV. The non-magnetic Ti$^{4+}$ ions contribute to lattice vibrations at all energies. 

Secondly, there is a considerable density of low-lying optical modes with energies as low as 5\,meV. These low-lying modes cross the longitudinal and transverse acoustic phonon branches, as can be seen in Fig.~\ref{fig:hto_phonons}, and appear in the phonon DOS, where they contribute substantially to the first peak, centered at 8\,meV, while the longitudinal acoustic phonon branches reach 14\,meV at the Brillouin zone boundaries. Recently, low-lying optical modes were identified in different rare earth pyrochlore materials, mainly zirconates, and it was established that their interference with the acoustic phonon modes could suppress the lattice thermal conductivity \cite{Lan:2015ba}. 

Thirdly, the phonon DOS contains a sharp spike at 15\,meV, dominated by the movement of rare earth ions. A sharp spike in the phonon DOS reflects nearly dispersionless phonon branches, which can be unambiguously distinguished in the dispersion relations presented in Fig.~\ref{fig:hto_phonons}. This result is particularly interesting for \tto{}~\cite{Princep:2015kt}, where there is a crystal field excitation with almost the same energy (we will discuss the consequences of this in a separate work~\cite{xtal_fields}).

\subsection{\label{sec:ins_cryst}Phonons from single crystal neutron spectroscopy}

\begin{figure}
  \centering
   \includegraphics[scale=1]{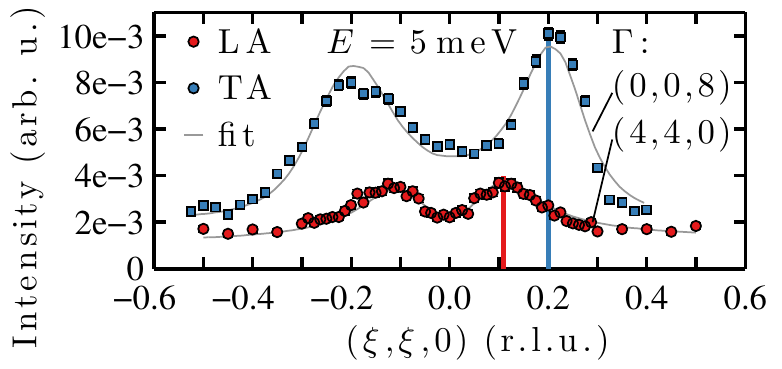}%
  \caption{Examples of constant energy scans across longitudinal acoustic (LA) and transverse acoustic (TA) phonon excitations propagating along the $(\zeta,\zeta,0)$ direction of the reciprocal lattice at an energy transfer of $5$\,meV at 200\,K, measured by INS. The LA mode was measured in the Brillouin zone of (4,4,0) and the TA mode at (0,0,8) to optimize the selection factor in the neutron cross section. The vertical solid lines indicate the fitted momentum transfers of the phonon excitations.}
   \label{fig:eiger_data}
\end{figure}

Guided by our DFT calculation, we expect the acoustic phonon branches in the energy range up to $E\simeq14$\,meV, which is conveniently accessible using thermal neutron triple axis spectroscopy. Exploiting the selection rule for nuclear excitations in the neutron scattering cross section, $(\vec{Q}\cdot\vec{e})^2$, where $\vec{Q}$ is the scattering vector and $\vec{e}$ is the polarization of the phonon, it is possible to choose Brillouin zones with high contrast between the acoustic phonon modes and the bundle of low-lying optical modes. The principle of the measurement and data analysis for the low-frequency phonon spectra is illustrated for acoustic phonons propagating along the high symmetry direction $(\zeta,\zeta,0)$ of the reciprocal lattice in Fig.~\ref{fig:eiger_data}.

The steep part of the phonon dispersion in the vicinity of the Brillouin zone center is best accessed with constant energy scans. Example scans collected on EIGER for \tto{} are presented in Figure \ref{fig:eiger_data}. Longitudinal phonon excitations $(\vec{e}=(1,1,0)/\sqrt{2})$ are probed in a longitudinal scan, that is $\vec{Q}\propto(h,h,0)$, in a Brillouin zone with a strong nuclear Bragg reflection, here $(4,4,0)$. The scan is symmetric with respect to the $\Gamma$ point. Transverse phonon excitations $(\vec{e}=(0,0,1))$, in contrast, are measured 
along $(h,h,8)$, since $(0,0,8)$ is the strongest accessible Bragg reflection along the $(0,0,l)$ direction. This scan has a focusing and defocusing side, which requires the consideration of the instrumental resolution for a proper description. All constant energy scans are fitted with an analytic dispersion model that approximates the acoustic phonon branches by an Arcus Tangent, which is a good approximation in the accessible momentum and energy space.  The dispersion model was convoluted with the instrumental resolution using the method of Popovici~\cite{POPOVICI:1975ww}.  Phonons at, or close to, the Brillouin zone boundaries have a vanishing slope and were therefore measured with constant-$\vec{Q}$ scans. Resolution effects in constant-$\vec{Q}$ scans across weakly or non-dispersing modes were neglected and the peak positions were determined from simple Gaussian fits.

\begin{figure}
  \centering
   \includegraphics[scale=1]{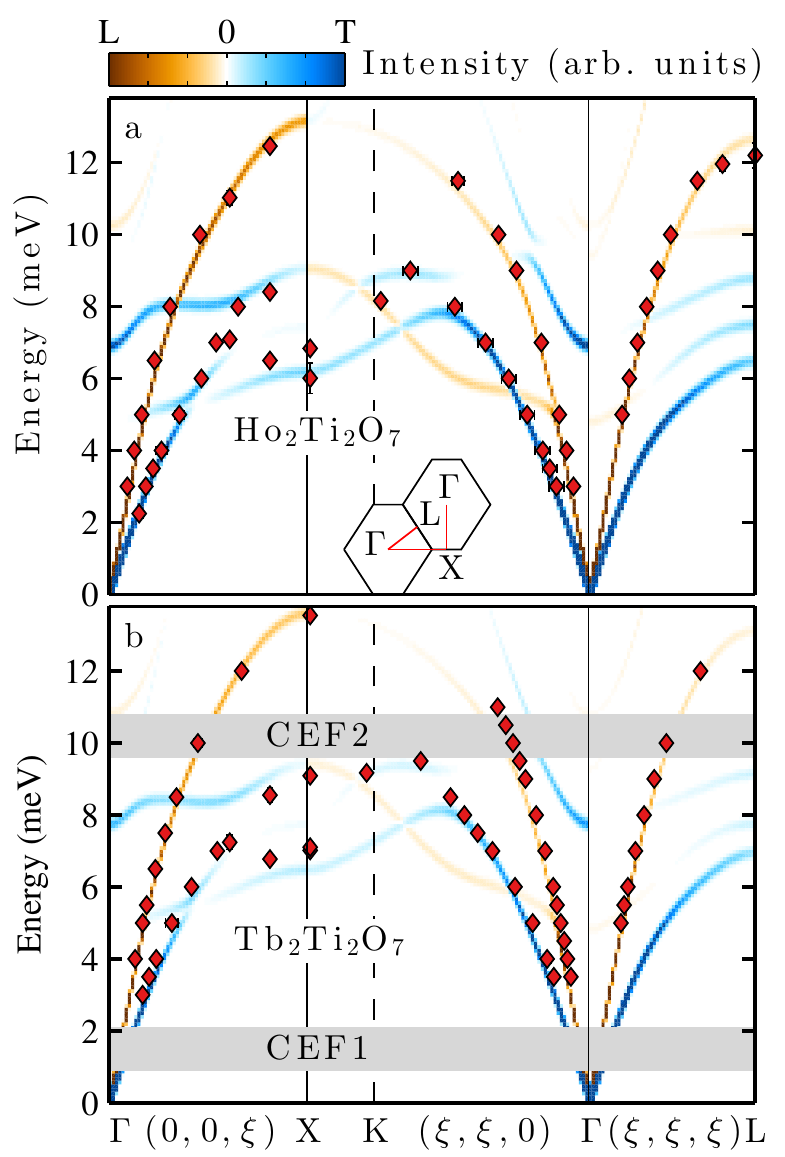}%
  \caption{Measurement of the acoustic phonon dispersion relations in \hto{} and \tto{} using INS. The color scale of the calculated spectrum shows the calculated intensities of longitudinal (L) and transverse (T) modes.  Within the constraints of our measurement geometry, we have detected almost all accessible modes, in their expected positions.  Fig.~\ref{fig:hto_phonons} shows that there are low-lying optical phonons within this energy range, but they have almost no measurable neutron scattering cross section so barely appear in this figure. The data are presented along a path of high symmetry directions in reciprocal space as indicated. In \tto{}, the measurement of phonon excitations was complicated by two strong ground state crystal electric field (CEF) excitations, shown schematically by the grey bands.}
   \label{fig:acoustic_phonons}
\end{figure}

Fig.~\ref{fig:acoustic_phonons} summarizes the extracted energies and momenta of phonon modes obtained from the TAS scans for \hto{} and \tto{}. The acoustic branches were measured in great detail and compare well with calculations. Due to the $\vec{Q}$-dependent polarization vectors of the phonon modes, the neutron intensity distribution along a given branch is non-trivial, but can be calculated from the theoretical phonon polarization vectors. In particular we find both theoretically and experimentally that the neutron cross section of the TA mode propagating along $(\zeta,\zeta,0)$ vanishes beyond the maximum of its dispersion. Note that the measurement geometry prevents access to the second TA mode, in which the ions vibrate orthogonal to the scattering plane. 

The different CEF splittings of the two rare earth ions Ho$^{3+}$ and Tb$^{3+}$ affect the accessibility of the acoustic phonon spectrum by neutron scattering in different ways. At $T=200$\,K, transitions between the thermally excited doublets at 22 and 26.5\,meV in \hto{} are possible.  However, their energy transfer is not larger than $\approx$ 4\,meV and their intensities are weak, and therefore do not perturb the measurement of the acoustic phonon branches. In contrast, in \tto{}, there are two intense ground state CEF transitions in the energy window of the acoustic modes, at 1.5\,meV (CEF1) and 10.2\,meV (CEF2). Constant-energy scans at energies close to these CEF excitations have a sloping background originating from the combination of rotating resolution ellipsoid and $\vec{Q}$-independent excitation.

\begin{figure}
  \centering
   \includegraphics[scale=1]{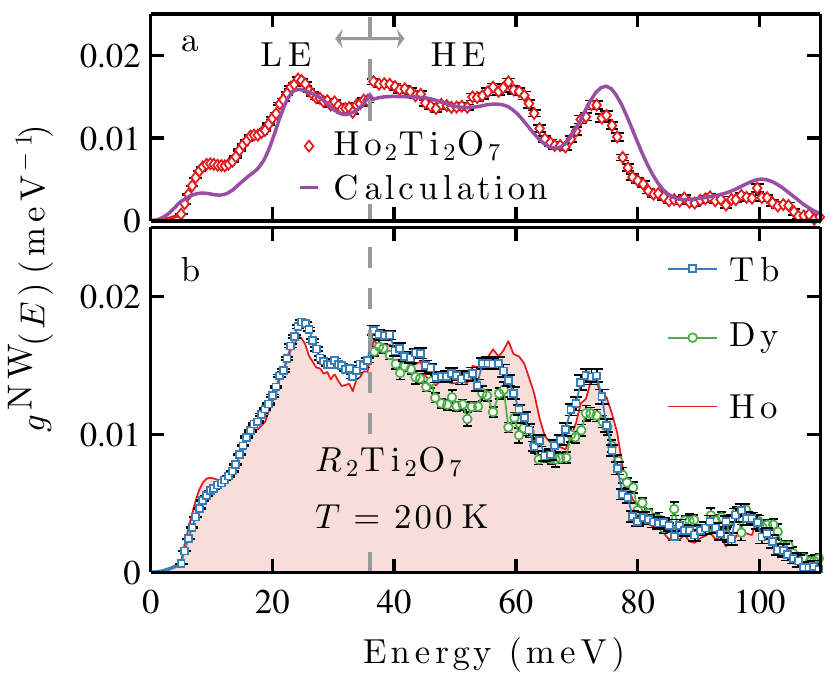}%
  \caption{Neutron-weighted phonon DOS $g^{\text{NW}}$ in the three rare earth titanate pyrochlores \tto{}, \dto{} and \hto{}. Panel a: Comparison of measured and calculated $g^{\text{NW}}$ for \hto{}. The phonon DOS is concatenated from low energy (LE) and high energy (HE) settings in order to optimize the resolution, as outlined in the main text. Panel b: Experimental $g^{\text{NW}}$ of \tto{} and \dto{} compared to \hto{} (colored patch). All measurements were performed at 200\,K.}
   \label{fig:exp_phdos}
\end{figure}

\subsection{\label{sec:ins_pow}Phonon DOS from powder neutron spectroscopy}

Using time-of-flight neutron spectroscopy we have measured the powder averaged excitations due to phonon scattering in all three rare earth titanates at $T=200$\,K, from which the neutron weighted phonon densities of states $g^{\rm NW}(E)$ can be extracted. Figure \ref{fig:exp_phdos} presents the experimental $g^{\rm NW}(E)$ along with the spectra obtained from our calculations.  The neutron-weighted phonon density of states was approximated as 
\begin{equation}
  g_{\text{NW}}(E)\simeq\sum_{d}\frac{\sigma_{d}}{m_d}g_{d}(E),
\end{equation}
where $g_{d}$ are the partial phonon densities of states for atom types $d$ (as shown for \hto{} in Fig.~\ref{fig:theory_pphdos}), $\sigma_{d}$ the total neutron scattering cross sections and $m_d$ the atomic masses \cite{NIST}. Here, the thermal Debye-Waller factor is approximated with unity, which is justified as the temperature of $T=200$\,K is low compared to the Debye temperature (estimated to be $\sim1100$ K~\cite{Pruneda:2005}), and the $|\vec{Q}|$ values of interest are relatively small. The calculated curve is convoluted with the elastic resolution functions of the respective time-of-flight setting and subsequently concatenated according to the experimental $g^{\rm NW}(E)$.  The contribution from the heavy rare earth ions which is prominent in the calculation shown in Fig.~\ref{fig:theory_pphdos} is strongly suppressed by the inverse mass term, which also promotes the oxygen dominated high energy part.  

The data from MERLIN, which has a much larger $|\vec{Q}|$-coverage, shows a similar phonon DOS for \hto{} and \tto{}, as expected from the single crystal data presented above.  Only the high energy modes are red-shifted noticeably in \tto{} compared to \hto{}.  The comparison of \dto{} is somewhat restricted since the lower $|\vec{Q}|$-coverage of 4SEASONS means that some energy ranges are contaminated by intense crystal field excitations and must be excluded, and averaging over Brillouin zones which contain significant phonon intensity is not as effective.  A useful comparison can be made in the high energy region, where we see that the phonon DOS of \dto{} is basically identical in form to \hto{} and \tto{}.  The theory reproduces all peak positions observed in the experimental phonon densities of states well, but not always their intensities. This disagreement may result from both the first-principles calculations, possible contributions from the phonon DOS of aluminum at $\sim 20$ meV and $\sim 35$ meV~\cite{Kresch:2008}, or the error introduced by the approximations applied to the experimental data. Based on the agreement of the theory with inelastic neutron and x-ray scattering from single crystals, however, we argue that the latter dominates.

\subsection{\label{sec:ixs}Phonons from x-ray spectroscopy}

Using IXS the optical phonon modes at higher energy transfer were studied. Our results demonstrate how complementary INS and IXS experiments can be used to confirm a large volume of the calculated phonon dispersion relations in \hto{}. Energy spectra were recorded at over 150 $\vec{Q}$-points in the $(h,h,l)$-plane, which were chosen carefully to optimize the contrast between neighboring phonon branches based on simulations of the calculated phonon spectrum. The measurements were carried out at different temperatures between 60\,K and room temperature. In this temperature range, energy shifts of phonon modes were not resolvable, but the peaks in the phonon spectra appeared broad at room temperature compared to temperatures well below, where the excitations became resolution limited. 

The combination of IXS measurements and DFT calculations of phonon spectra enables a two-fold analysis of our data: on one hand, the fitting of phonon excitations in the inelastic energy scans, and on the other hand the calculation of inelastic x-ray intensities from \textit{ab initio} eigenvalues and eigenvectors of the dynamical matrices. In the first approach, all $\vec{Q}$-points that fall in the reduced unit cell within an interval of $\pm0.05$ reduced lattice units around one of the high symmetry directions are considered for fitting. The phonon excitations are described by Lorentzian functions that are convoluted with the pseudovoigt resolution function of the respective analyzer. The elastic line is described by a pseudovoigt function with the known analyzer-dependent parameters and variable amplitude.  Fig.~\ref{fig:hto_phonons} shows all phonon dispersion points extracted from IXS scans along high symmetry directions, combined with the single crystal neutron scattering results, compared with the phonon dispersions computed by density functional theory. 

\begin{figure}
  \centering
   \includegraphics[scale=1]{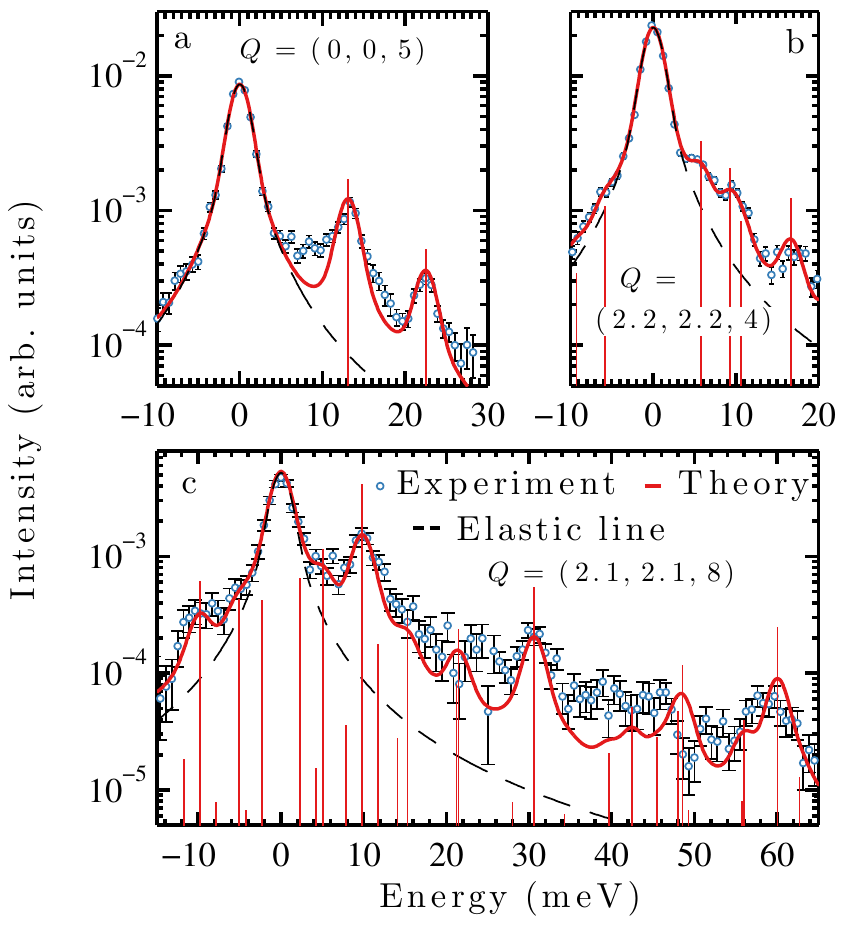}%
  \caption{Experimental and theoretical IXS intensities at selected $\vec{Q}$-points. The calculated IXS intensity (indicated by vertical lines, scaled) was convoluted with the instrumental resolution and an elastic line was added (dashed line). For each spectrum the theoretical curves were scaled to the experimental data by a constant factor to account for sample attenuation.}
   \label{fig:exp_vs_theory_ixs}
\end{figure}

In the second approach, we directly compare the measured energy scans to the calculated IXS intensity from the \textit{ab initio} result using the software ab2tds~\cite{Ab2tds}. The calculated intensities are convoluted with the instrumental resolution, the elastic line is added and a global intensity scaling due to absorption of the sample is applied. The analysis has been applied to all measured $\vec{Q}$-points and we do not find any systematic deviation of the experimental data from the theory. In Fig.~\ref{fig:exp_vs_theory_ixs}, we highlight the agreement between experimental and simulated IXS scans for selected $\vec{Q}$-points over a wide range of energy transfers.  

\begin{figure}
  \centering
   \includegraphics[scale=1]{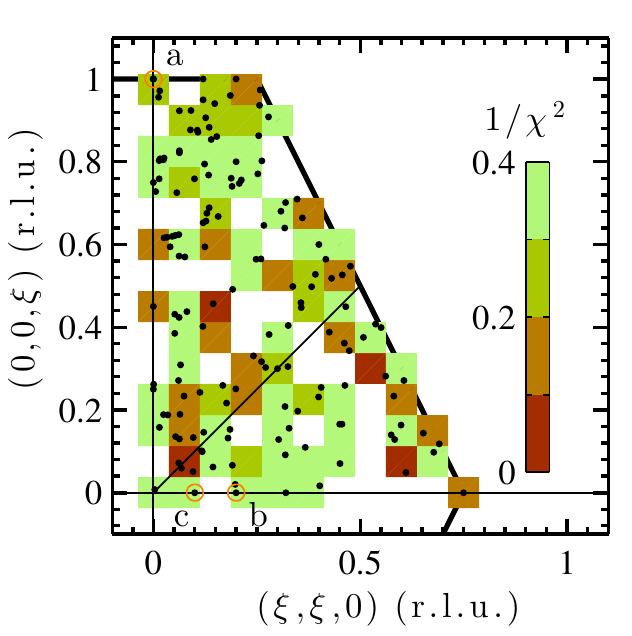}%
  \caption{Goodness of calculation determined at every $\vec{Q}$-point measured and represented in the symmetry reduced Brillouin zone. For each spectrum (denoted by black dots), $\xi^2$ is calculated as the standard deviation of the theoretical curve (convoluted with instrumental resolution, elastic line added) from the experimental data (the values of $1/\chi^2$ falling within the small square tiles are then binned and used to encode its color).  The points of the individual scans shown in Fig.~\ref{fig:exp_vs_theory_ixs} a, b, and c are labelled and marked by rings.}
   \label{fig:goodness_of_calc}
\end{figure}

In Fig.~\ref{fig:goodness_of_calc} we present the goodness of the first-principles phonon calculation which is computed as the inverse mean square deviation of the simulated intensity from the measured x-ray intensities at all $\vec{Q}$-points.  The figure provides insight into the agreement between calculation and experiment not only along the high symmetry lines, but also at general $\vec{Q}$ in the reduced unit cell. Note that because the majority of the spectra were recorded up to $E\approx25-30$\,meV, the goodness of the calculation is not comparable for higher energies. At individual $\vec{Q}$-points there may be deviations of the calculated spectrum from the measurement (in both energy and intensity), which can reach from a single mode up to the entire spectrum. While severe mismatches often involve the presence (absence) of phonon excitations around 10\,meV and/or 20-25\,meV in energy space, these are not systematic in $\vec{Q}$-space. In particular we can rule out correlations between deviations and $\vec{Q}$-points as well as the analyzer (detector) channels, and find that considerable disagreement between calculation and experiment appear in less than 10\% of all measured $\vec{Q}$ positions. The level of agreement presented in Figure.~\ref{fig:exp_vs_theory_ixs} is representative of approximately 60\% of all measured spectra.  Since the IXS phonon intensity depends sensitively on the eigenvectors, it is noteworthy that the intensity ratios of the computed phonon excitations are generally in good agreement with the experiment.

\section{\label{sec:Discussion}Discussion}

In the preceding sections we have presented a model of the lattice dynamics in idealized rare earth titanate pyrochlores, which allows the calculation of the phonon dispersion relations and the assignment of the symmetries of the modes.  Along with this model calculation, we have performed various neutron and x-ray scattering experiments which we have used to test, or validate, the description of lattice dynamics of rare earth titanates by this model.  Generally, we have obtained a close agreement between experimentally determined phonon frequencies and the calculated dispersion relations, as can be seen in Fig.~\ref{fig:hto_phonons} and Fig.~\ref{fig:acoustic_phonons}, for example.  Our IXS study probes not only the energies, but also the eigenvectors of the dynamical matrix, and the close agreement of both calculated energies and intensities, as shown in Fig.~\ref{fig:exp_vs_theory_ixs} and Fig.~\ref{fig:goodness_of_calc}, shows that these eigenvectors are also realistic.  

The model we have employed has the 4$f$ electrons of the rare earth ions frozen in the core, so, at the level of these calculations, the only differences between the rare earth ions are their mass and ionic radius.  We have therefore calculated the end members of our series ($r_{\mathrm{Tb}^{3+}}>r_{\mathrm{Dy}^{3+}}>r_{\mathrm{Ho}^{3+}}$) and see that the resulting differences in phonon energies are tiny, as can be seen in the comparison of acoustic modes shown in Fig.~\ref{fig:acoustic_phonons}.  This is borne out by the comparison of the experimental phonon DOS in all three compounds (Fig.~\ref{fig:exp_phdos}).  Results such as Fig.~\ref{fig:hto_phonons} and Table~\ref{tab:phonon_symmetries} can therefore be regarded as a good guide to the phonon spectrum of all the heavy rare earth titanates - modes of the same symmetry and closely similar energy can be expected across the entire series from $R =$ Gd to $R =$ Yb, with progressive shift to slightly higher energies as the ionic radius contracts.  As more and more studies seek exotic excitations in these materials, using all types of spectroscopy, such a guide is useful in clarifying the assignment of signals, as we will discuss further below.

The evaluation of non-analytical term corrections reveal that all $F_{1u}$ modes are polar and split into doubly degenerate $E$-modes (TO) and non-degenerate $A_1$ modes (LO). To the best of our knowledge, this is the first application of non-analytical term corrections to the dynamical matrix of rare earth titanates. The resulting LO/TO splitting is best distinguished in comparison with Ref.~[\onlinecite{Kumar:2012}], where it is not included, and is supported by our IXS measurements.  The splitting is particularly important for the optical phonon at $\sim15$ meV, as seen explicitly in the dispersion relations of Fig.~\ref{fig:hto_phonons} in the vicinity of the second $\Gamma$-point.  There, the triply degenerate $F_{1u}$ mode has been split and we calculate a weakly dispersive phonon mode at $\sim15$ meV, and a new mode phonon mode at $\sim17$ meV.  The calculation shows that the mode at $\sim 15$ meV is doubly degenerate at the $\Gamma$-point, and hence is the TO-modes, while the mode at $\sim17$ meV is non-degenerate at the $\Gamma$-point, and hence is the LO-mode.  Their existence, energies and eigenvectors were all confirmed by comparison with IXS measurements at finite $\vec{Q}$.

\begin{figure}
  \centering
   \includegraphics[scale=0.95]{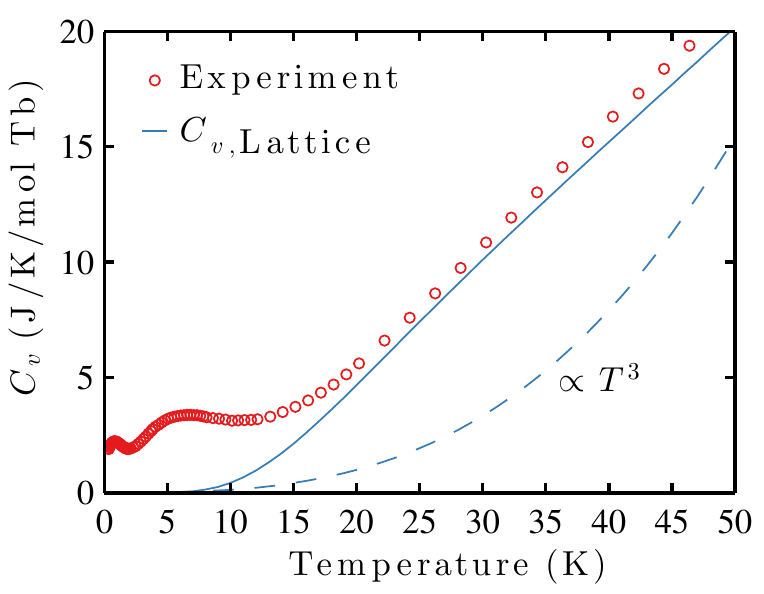}
  \caption{Comparison of the experimental specific heat of a \tto{} single crystal, as originally presented in Ref.~[\onlinecite{Fennell:2014gf} (supplementary material)] and discussed further in Ref.~[\onlinecite{mem_all_samps}], with the lattice contribution derived from the density functional calculations presented here.  We suggest that DFT calculations could provide a means to accurately estimate the lattice contribution to such measurements.}
   \label{fig:lattice_cvs}
\end{figure}

Our density functional calculations are performed at $T=0$ K.  However, with the knowledge of the phonon band structure, an array of thermodynamic quantities are directly accessible, such as the free energy, heat capacity, or the entropy. Particularly interesting with respect to the frustrated magnetism in spin ice materials and \tto{} is the lattice contribution to the specific heat.  Experimentally, it is challenging to separate the magnetic contribution to the specific heat.  For example, in \tto{} the low energy crystal field excitations and strong magnetic fluctuations contribute over a wide temperature range, making the simple parameterization of the lattice contribution difficult, frustrating attempts to definitively estimate the magnetic entropy~\cite{Chapuis:2010ir}.  Since our calculation is designed to be non-magnetic (because the the 4$f$ electrons are frozen into the core states), it separates the lattice contribution to the specific heat directly. We show the example of \tto{} in Fig.~\ref{fig:lattice_cvs}.  The calculated $C_{v, \textrm{Lattice}}$ fits well to the measurement, but, due to the contribution from many relatively low-lying optical modes, it does not resemble a simple $T^3$ law, except at the very lowest temperatures.  For reference, the calculated lattice specific heat is tabulated for \hto{} and \tto{} in Appendix~\ref{appendix:lattice_cvs}.  Lattice heat capacities have not been previously estimated for heat capacity measurements of rare earth titanates in this way, and could be used to discriminate purely magnetic contributions to $C_v$.

Also relating to the effect of finite temperatures are other important lattice dynamical properties which are not captured by our calculations, nor pursued in our experiments.  Nonlinear anharmonic effects and phonon-phonon scattering may occur at higher temperature, and these would have to be investigated for a complete understanding of the lattice dynamics of the \rto{} materials.  We noted in passing that in the IXS experiment, the peak width of many optical phonons was considerably broadened at room temperature, and this could be taken as a sign that such effects do indeed occur.  Similarly, the effect of pressure variation may also be interesting.

\begin{figure}
  \centering
   \includegraphics[scale=1]{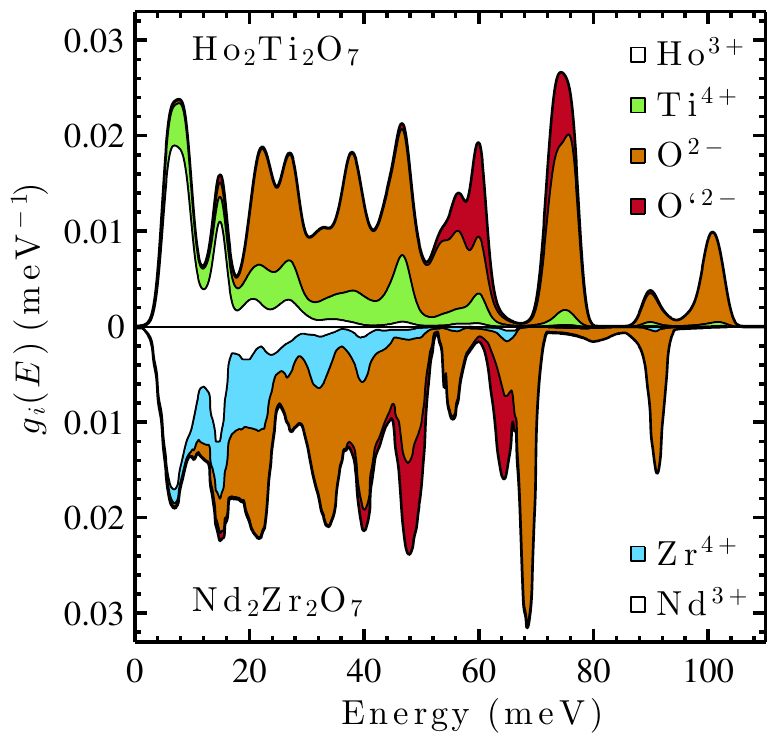}%
  \caption{Comparison of the calculated phonon partial DOS in \hto{} and Nd$_2$Zr$_2$O$_7$ (taken from Ref.~[\onlinecite{Lan:2015ba}]) exemplifying two trends among different rare earth pyrochlores: general shift to lower energy with lattice expansion; and important shift of the partial phonon DOS of modes involving the B cation to lower energy with increasing mass of that cation.  The partial phonon DOS of modes involving the rare earth ion are very similar due to the relatively small change in mass among different rare earth ions in different pyrochlore materials.}
   \label{fig:hto_nzo_comparison}
\end{figure}

Our calculations of the pyrochlore titanate lattice can also serve as a useful guide to lattice vibrations in other rare earth pyrochlore materials with tetrapositive B-site ions, such as zirconate or hafnate pyrochlores. However, according to the phonon calculations presented in Ref.~[\onlinecite{Lan:2015ba}], pyrochlores in which the B-site cation is a member of group IV, such as Sn, Ge or Pb, have significantly different phonon frequencies.  In comparison with the partial phonon densities of states calculated for a subset of zirconate and hafnate pyrochlores~\cite{Lan:2015ba}, we find that the phonon band structures of (3+,4+) pyrochlores have essentially identical features, with adjustments that can be classified generally in two ways, as illustrated in Fig.~\ref{fig:hto_nzo_comparison}: Firstly, the larger ionic radii of both A-and B-ions lead to an expansion of the unit cell, which reduces the frequencies of phonon vibrations across the entire phonon spectrum, especially of the phonon modes dominated by the light oxygen ions. Secondly, the mass of the B-ions changes dramatically between Ti, Zr, and Hf, by factors of 1.9, and 3.7 with respect to the mass of Ti. The larger the mass of the B-ion, the more the statistical weight of its partial phonon density of states will be shifted to lower frequencies compared to $g_{\rm Ti}(E)$ in \rto{}. Since the mass of the heavy rare earth ions changes only marginally along the lanthanide series, its partial phonon density of states has a comparable distribution in all rare earth pyrochlores.

The present calculation was carried out for a squeezed unit cell to stabilize otherwise imaginary phonon modes. This procedure is justified by the extensive verification of the calculated phonon spectrum using INS and IXS techniques.  We also find a very close agreement between the phonon modes at the $\Gamma$-point in our work and the LDA calculation of Kumar \textit{et al.}~\cite{Kumar:2012}. Since the $4f$ electrons were frozen in the core, magnetism is excluded from the present calculation. By including the $4f$ electrons along with a Hubbard $U$ parameter, or by the use of hybrid functionals, one may hope to describe magnetoelastic effects observed experimentally in rare earth titante pyrochlores, especially for $R=$ Tb.  Treating the 4$f$ electrons explicitly in the DFT calculation may not lead to significant changes in the phonon dispersion calculated at $T=0$ K \cite{Chernyshev:2015}, but could provide a qualitative description of magnetoelastic effects in the thermodynamic properties of the three compounds, especially in \tto{}, which were observed experimentally \cite{Nakanishi:2011bz}. The theoretical treatment of coupling  of the lattice dynamics with the spin system may require an effective model Hamiltonian acting on a subset of the combined crystal field and phonon phase space, mapping out only the relevant crystal field states and normal modes of vibration. Our calculation is useful in this respect, as allows us to identify possible interactions of vibrational modes with the known crystal field spectra of rare earth titanate,  zirconate and hafnate pyrochlores.

\section{\label{sec:Conclusion}Conclusion}

The phonon spectra of the rare earth titanate pyrochlores \tto{} and \hto{} were calculated using density functional theory.  The symmetries, energies and eigenvectors obtained from this model were thoroughly verified by a combination of inelastic neutron and x-ray scattering on \tto{} and \hto{}.  Comparison of the phonon DOS of \dto{} with those of \hto{} and \tto{} shows that it also has a very similar phonon band structure.  The resulting reliable energies, eigenvectors and zone center symmetries of the vibrational modes (as tabulated in Table~\ref{tab:phonon_symmetries}), dispersion relations (as shown in Fig.~\ref{fig:hto_phonons} and Fig.~\ref{fig:acoustic_phonons}), and densities of states (Fig.~\ref{fig:theory_pphdos}),  are important guides for the study of magnetoelastic effects and other excitations in \rto{}.  General properties of the lattice dynamics of rare earth pyrochlores with various B-site cations were pointed out in comparison to those studied in this work.

\acknowledgements{

MR acknowledges constructive discussions with M. Troyer (ETH Zurich) and P. Blaha (TU Vienna), we thank S. T. Bramwell (UCL) for encouraging us to calculate the lattice heat capacities.  The inelastic x-ray experiments were performed on beamline ID28 at the European Synchrotron Radiation Facility (ESRF), Grenoble, France.  Neutron scattering experiments were carried out at ISIS (Rutherford Appleton Laboratory, UK), MLF (J-PARC, Japan), and the continuous spallation neutron source SINQ at the Paul Scherrer Institut at Villigen PSI (Switzerland); x-ray diffraction was carried out at the Swiss Light Source.  Computations were performed on the Brutus cluster at ETH Zurich.  Work at PSI was partly funded by the SNSF (Schweizerischer Nationalfonds zur F\"orderung der Wissenschaftlichen Forschung) (grant 200021\_140862 and 200020\_162626), MN and NAS were supported by the ETH Zurich and the ERC Advanced Grant program, No. 291151, BW acknowledges financial support from the European Community's Seventh Framework Programme (FP7/2007-2013) under grant agreement n.$^\mathrm{o}$290605 (PSI-FELLOW/COFUND), work done in Oxford was funded by the EPSRC, UK (grants EP/K028960/1 and EP/M020517/1), LB is supported by The Leverhulme Trust through the Early Career Fellowship program.}

\appendix

\section{Characterization of powder samples}\label{appendix:characterization}

The structures of polycrystalline samples of \hto{}, \dto{} and \tto{} were verified by powder neutron and x-ray diffraction experiments performed at HRPT~\cite{Fischer:2000to} (at SINQ, PSI) and the MSB (Materials Science Beamline)~\cite{Willmott:2013gv} (at the SLS, PSI), respectively. All powder diffraction measurements were carried out at room temperature. For the neutron measurements the powder samples were filled in vanadium cylinders, mounted directly on the sample table of the diffractometer, and measured with incoming neutron wave-length $\lambda=1.15$\,\AA ~for 4 hours ($R=$ Ho, Tb) or 7 hours ($R=$ Dy).   The MSB was operated in Debye-Scherrer geometry with the Mythen microstrip detector, capillary spinner, and $2\theta$ range extending from $2^\mathrm{\circ}$ to 120$^\mathrm{\circ}$.  The x-ray wavelength was $\lambda=0.4959$~\AA~(i.e. $E=25$ keV). The powder samples were diluted with the Si standard NIST640C and filled into a 0.5 mm diameter glass capillary, mounted and aligned on the diffractometer, measuring time was 16 minutes per sample.

In Table.~\ref{tab:TTO_refinement} we present the crystallographic parameters of the powder samples described above, as determined by joint Rietveld refinements against their powder neutron and x-ray diffraction patterns.  Based on the wavelength calibration using the silicon standard (mixed with the sample) the lattice parameters of \rto{} were refined accurately from the x-ray patterns. The structures of the three \rto{} samples were entirely refined from the neutron patterns. In the refinement the occupation of Ti$^{4+}$ ions was fixed to 1.0 and the relative fractional occupations of the rare earth and the two oxygen sites were determined. At the 1\% level no deviations from stoichiometry were found in any of the three powder samples. Due to the strong absorption of natural isotopic abundance dysprosium, the anisotropic thermal displacement factors of \dto{} are somewhat less reliable in comparison with \hto{} and \tto{}.

\begin{table}
\ra{1.3}
\centering
\caption{Crystallographic parameters of \rto{}, space group $Fd\bar{3}m$, as determined from joint Rietveld refinements of x-ray and neutron diffraction data.}
\begin{tabular}{l c c c c c c c c c c c c}
\hline
\hline
\multicolumn{13}{l}{\tto{}} \\
\multicolumn{8}{l}{$a=10.15291(1)$\,(\AA)} & \multicolumn{5}{r}{R$_{\rm Bragg}=2.33$} \\ \hline
\multicolumn{13}{l}{Atomic coordinates} \\ \hline
\multicolumn{2}{l}{Atom}   & \multicolumn{3}{c}{$x$}       & \multicolumn{3}{c}{$y$}   & \multicolumn{3}{c}{$z$}  & \multicolumn{2}{r}{Frac. occ.}   \\ 
\multicolumn{2}{l}{Tb(16d)}& \multicolumn{3}{c}{0.5000}    & \multicolumn{3}{c}{0.5000}& \multicolumn{3}{c}{0.5000}& \multicolumn{2}{r}{1.001(25)} \\
\multicolumn{2}{l}{Ti(16c)}& \multicolumn{3}{c}{0.0000}    & \multicolumn{3}{c}{0.0000}& \multicolumn{3}{c}{0.0000}& \multicolumn{2}{r}{1.000} \\
\multicolumn{2}{l}{O(48f)} & \multicolumn{3}{c}{0.3279(2)} & \multicolumn{3}{c}{0.1250}& \multicolumn{3}{c}{0.1250}& \multicolumn{2}{r}{0.997(12)} \\
\multicolumn{2}{l}{O(8b)}  & \multicolumn{3}{c}{0.3750}    & \multicolumn{3}{c}{0.3750}& \multicolumn{3}{c}{0.3750}& \multicolumn{2}{r}{0.997(24)} \\ \hline
\multicolumn{13}{l}{Anisotropic displacement parameters (\AA$^2\times10^4$)} \\ \hline
       & \multicolumn{2}{c}{B$_{11}$} & \multicolumn{2}{c}{B$_{22}$} & \multicolumn{2}{c}{B$_{33}$} & \multicolumn{2}{c}{B$_{12}$} & \multicolumn{2}{c}{B$_{13}$} & \multicolumn{2}{c}{B$_{23}$} \\ 
Tb(16d)& \multicolumn{2}{c}{13.7(8)} & \multicolumn{2}{c}{13.7(8)} & \multicolumn{2}{c}{13.7(8)} & \multicolumn{2}{c}{-4.8(6)} & \multicolumn{2}{c}{-4.8(6)} & \multicolumn{2}{c}{-4.8(6)}   \\ 
Ti(16c)& \multicolumn{2}{c}{11.1(11)} & \multicolumn{2}{c}{11.1(11)} & \multicolumn{2}{c}{11.1(11)} & \multicolumn{2}{c}{1.6(15)} & \multicolumn{2}{c}{1.6(15)} & \multicolumn{2}{c}{1.6(15)} \\
O(48f) & \multicolumn{2}{c}{13.5(12)} & \multicolumn{2}{c}{10.3(7)} & \multicolumn{2}{c}{10.3(7)} & \multicolumn{2}{c}{0.0} & \multicolumn{2}{c}{0.0} & \multicolumn{2}{c}{3.8(10)}        \\
O(8b)  & \multicolumn{2}{c}{9.2(14)} & \multicolumn{2}{c}{9.2(14)} & \multicolumn{2}{c}{9.2(14)} & \multicolumn{2}{c}{0.0} & \multicolumn{2}{c}{0.0} & \multicolumn{2}{c}{0.0}               \\
\hline\hline

\multicolumn{13}{l}{\dto{}} \\
\multicolumn{8}{l}{$a=10.12523(1)$\,(\AA)} & \multicolumn{5}{r}{R$_{\rm Bragg}=2.98$} \\ \hline
\multicolumn{13}{l}{Atomic coordinates} \\ \hline
\multicolumn{2}{l}{Atom}   & \multicolumn{3}{c}{$x$}       & \multicolumn{3}{c}{$y$}  & \multicolumn{3}{c}{$z$}  & \multicolumn{2}{r}{Frac. occ.}   \\
\multicolumn{2}{l}{Dy(16d)}& \multicolumn{3}{c}{0.5000}    & \multicolumn{3}{c}{0.5000}& \multicolumn{3}{c}{0.5000}& \multicolumn{2}{r}{1.046(72)} \\
\multicolumn{2}{l}{Ti(16c)}& \multicolumn{3}{c}{0.0000}    & \multicolumn{3}{c}{0.0000}& \multicolumn{3}{c}{0.0000}& \multicolumn{2}{r}{1.000} \\
\multicolumn{2}{l}{O(48f)} & \multicolumn{3}{c}{0.3287(6)} & \multicolumn{3}{c}{0.1250}& \multicolumn{3}{c}{0.1250}& \multicolumn{2}{r}{1.041(72)} \\
\multicolumn{2}{l}{O(8b)}  & \multicolumn{3}{c}{0.3750}    & \multicolumn{3}{c}{0.3750}& \multicolumn{3}{c}{0.3750}& \multicolumn{2}{r}{1.007(94)} \\ \hline
\multicolumn{13}{l}{Anisotropic displacement parameters (\AA$^2\times10^4$)} \\ \hline
       & \multicolumn{2}{c}{B$_{11}$} & \multicolumn{2}{c}{B$_{22}$} & \multicolumn{2}{c}{B$_{33}$} & \multicolumn{2}{c}{B$_{12}$} & \multicolumn{2}{c}{B$_{13}$} & \multicolumn{2}{c}{B$_{23}$} \\  
Dy(16d)& \multicolumn{2}{c}{4.3(10)} & \multicolumn{2}{c}{4.3(10)} & \multicolumn{2}{c}{4.3(10)} & \multicolumn{2}{c}{2.5(54)} & \multicolumn{2}{c}{2.5(54)} & \multicolumn{2}{c}{2.5(54)}   \\ 
Ti(16c)& \multicolumn{2}{c}{-0.9(40)} & \multicolumn{2}{c}{-0.9(40)} & \multicolumn{2}{c}{-0.9(40)} & \multicolumn{2}{c}{1.6(15)} & \multicolumn{2}{c}{1.6(15)} & \multicolumn{2}{c}{1.6(15)} \\
O(48f) & \multicolumn{2}{c}{5.1(38)} & \multicolumn{2}{c}{0.7(25)} & \multicolumn{2}{c}{0.7(25)} & \multicolumn{2}{c}{0.0} & \multicolumn{2}{c}{0.0} & \multicolumn{2}{c}{2.6(30)}        \\
O(8b)  & \multicolumn{2}{c}{-1.2(40)} & \multicolumn{2}{c}{-1.2(40)} & \multicolumn{2}{c}{1-2.(40)} & \multicolumn{2}{c}{0.0} & \multicolumn{2}{c}{0.0} & \multicolumn{2}{c}{0.0}               \\
\hline\hline
\multicolumn{13}{l}{\hto{}}  \\
\multicolumn{8}{l}{$a=10.10186(1)$\,(\AA)} & \multicolumn{5}{r}{R$_{\rm Bragg}=2.74$} \\ \hline
\multicolumn{13}{l}{Atomic coordinates} \\ \hline
\multicolumn{2}{l}{Atom}   & \multicolumn{3}{c}{$x$}       & \multicolumn{3}{c}{$y$}   & \multicolumn{3}{c}{$z$}   & \multicolumn{2}{r}{Frac. occ.}   \\
\multicolumn{2}{l}{Ho(16d)}& \multicolumn{3}{c}{0.5000}    & \multicolumn{3}{c}{0.5000}& \multicolumn{3}{c}{0.5000}& \multicolumn{2}{r}{1.047(24)} \\
\multicolumn{2}{l}{Ti(16c)}& \multicolumn{3}{c}{0.0000}    & \multicolumn{3}{c}{0.0000}& \multicolumn{3}{c}{0.0000}& \multicolumn{2}{r}{1.000} \\
\multicolumn{2}{l}{O(48f)} & \multicolumn{3}{c}{0.3293(2)} & \multicolumn{3}{c}{0.1250}& \multicolumn{3}{c}{0.1250}& \multicolumn{2}{r}{1.006(23)} \\
\multicolumn{2}{l}{O(8b)}  & \multicolumn{3}{c}{0.3750}    & \multicolumn{3}{c}{0.3750}& \multicolumn{3}{c}{0.3750}& \multicolumn{2}{r}{0.997(30)} \\ \hline
\multicolumn{13}{l}{Anisotropic displacement parameters (\AA$^2\times10^4$)} \\ \hline
       & \multicolumn{2}{c}{B$_{11}$} & \multicolumn{2}{c}{B$_{22}$} & \multicolumn{2}{c}{B$_{33}$} & \multicolumn{2}{c}{B$_{12}$} & \multicolumn{2}{c}{B$_{13}$} & \multicolumn{2}{c}{B$_{23}$} \\ 
Ho(16d)& \multicolumn{2}{c}{14.2(7)} & \multicolumn{2}{c}{14.2(7)} & \multicolumn{2}{c}{14.2(7)} & \multicolumn{2}{c}{-4.3(6)} & \multicolumn{2}{c}{-4.3(6)} & \multicolumn{2}{c}{-4.3(6)}   \\ 
Ti(16c)& \multicolumn{2}{c}{9.9(15)} & \multicolumn{2}{c}{9.9(15)} & \multicolumn{2}{c}{9.9(15)} & \multicolumn{2}{c}{-0.9(14)} & \multicolumn{2}{c}{-0.9(14)} & \multicolumn{2}{c}{-0.9(14)} \\
O(48f) & \multicolumn{2}{c}{15.0(12)} & \multicolumn{2}{c}{9.9(6)} & \multicolumn{2}{c}{9.9(6)} & \multicolumn{2}{c}{0.0} & \multicolumn{2}{c}{0.0} & \multicolumn{2}{c}{3.8(9)}        \\
O(8b)  & \multicolumn{2}{c}{7.4(13)} & \multicolumn{2}{c}{7.4(13)} & \multicolumn{2}{c}{7.4(13)} & \multicolumn{2}{c}{0.0} & \multicolumn{2}{c}{0.0} & \multicolumn{2}{c}{0.0}               \\
\hline
\hline

\end{tabular}
\label{tab:TTO_refinement}
\end{table}

\section{Lattice heat capacities}\label{appendix:lattice_cvs}

In Tables~\ref{tab:TTO_C_v} and ~\ref{tab:HTO_C_v}, we tabulate the calculated lattice heat capacities of \tto{} and \hto{} respectively.

\begin{table}
\caption{Calculated lattice contribution to the specific heat $C_v$ of \tto{} in units of J/K/mol Tb as function of temperature. The values correspond to the line in Fig.~\ref{fig:lattice_cvs}. The  sampling mesh for the specific heat is carefully chosen to increase the convergence at low temperatures. With a mesh of size 71x71x71 we yield the following convergence: At $T=2$\,K $4\times10^{-2}$, at 10\,K  $4\times10^{-5}$ and above 100\,K better than $2\times10^{-7}$.}
\ra{1.3}
\centering
\begin{tabular}{r c c r r c r r}\hline\hline
\multicolumn{1}{c}{$T$\,(K)} &  $C_v$\,(J/K/mol Tb)  & &  \multicolumn{1}{c}{$T$}  &  \multicolumn{1}{c}{$C_v$}  & &  \multicolumn{1}{c}{$T$}  &  \multicolumn{1}{c}{$C_v$} \\
 \hline
0.0 & 0.0000 & & 22.0 & 5.7569 & & 120.0  & 54.8169 \\
2.0 & 0.0004 & & 24.0 & 6.8569 & & 140.0  & 63.5991 \\
3.0 & 0.0015 & & 26.0 & 7.9489 & & 160.0  & 71.5477 \\
4.0 & 0.0039 & & 28.0 & 9.0267 & & 180.0  & 78.6392 \\
5.0 & 0.0096 & & 30.0 & 10.0880 & & 200.0  & 84.9057 \\
6.0 & 0.0248 & & 32.0 & 11.1333 & & 225.0  & 91.6765 \\
7.0 & 0.0610 & & 34.0 & 12.1643 & & 250.0  & 97.4090 \\
8.0 & 0.1320 & & 36.0 & 13.1834 & & 275.0  & 102.2572 \\
9.0 & 0.2504 & & 38.0 & 14.1932 & & 300.0  & 106.3633 \\
10.0 & 0.4242 & & 40.0 & 15.1962 & & 350.0  & 112.8248 \\
11.0 & 0.6571 & & 45.0 & 17.6876 & & 400.0  & 117.5636 \\
12.0 & 0.9481 & & 50.0 & 20.1748 & & 450.0  & 121.1055 \\
13.0 & 1.2930 & & 55.0 & 22.6714 & & 500.0  & 123.8039 \\
14.0 & 1.6857 & & 60.0 & 25.1823 & & 550.0  & 125.8971 \\
15.0 & 2.1191 & & 65.0 & 27.7068 & & 600.0  & 127.5480 \\
16.0 & 2.5861 & & 70.0 & 30.2410 & & 650.0  & 128.8698 \\
17.0 & 3.0797 & & 75.0 & 32.7793 & & 700.0  & 129.9426 \\
18.0 & 3.5937 & & 80.0 & 35.3155 & & 800.0  & 131.5566 \\
19.0 & 4.1227 & & 90.0 & 40.3555 & & 900.0  & 132.6917 \\
20.0 & 4.6619 & & 100.0 & 45.3121 & & 1000.0  & 133.5184 \\
\hline\hline
\end{tabular}
\label{tab:TTO_C_v}
\end{table}

\begin{table}
\caption{Calculated lattice contribution to the specific heat $C_v$ of \hto{} in units of J/K/mol Ho as function of temperature. The sampling mesh for the specific heat is identical to that used for \tto{} (see caption of Table \ref{tab:TTO_C_v}) and yields the same degree of convergence.}
\ra{1.3}
\centering
\begin{tabular}{r c c r r c r r}\hline\hline
\multicolumn{1}{c}{$T$\,(K)} &  $C_v$\,(J/K/mol Ho)  & &  \multicolumn{1}{c}{$T$}  &  \multicolumn{1}{c}{$C_v$}  & &  \multicolumn{1}{c}{$T$}  &  \multicolumn{1}{c}{$C_v$} \\
 \hline
0.0 & 0.0000 & & 22.0 & 6.2627 & & 120.0  & 55.1878 \\
2.0 & 0.0004 & & 24.0 & 7.3972 & & 140.0  & 63.8150 \\
3.0 & 0.0015 & & 26.0 & 8.5173 & & 160.0  & 71.6385 \\
4.0 & 0.0041 & & 28.0 & 9.6187 & & 180.0  & 78.6347 \\
5.0 & 0.0109 & & 30.0 & 10.7008 & & 200.0  & 84.8316 \\
6.0 & 0.0303 & & 32.0 & 11.7651 & & 225.0  & 91.5443 \\
7.0 & 0.0763 & & 34.0 & 12.8141 & & 250.0  & 97.2421 \\
8.0 & 0.1651 & & 36.0 & 13.8505 & & 275.0  & 102.0719 \\
9.0 & 0.3094 & & 38.0 & 14.8769 & & 300.0  & 106.1704 \\
10.0 & 0.5166 & & 40.0 & 15.8959 & & 350.0  & 112.6351 \\
11.0 & 0.7882 & & 45.0 & 18.4229 & & 400.0  & 117.3885 \\
12.0 & 1.1212 & & 50.0 & 20.9373 & & 450.0  & 120.9483 \\
13.0 & 1.5095 & & 55.0 & 23.4504 & & 500.0  & 123.6642 \\
14.0 & 1.9451 & & 60.0 & 25.9664 & & 550.0  & 125.7734 \\
15.0 & 2.4198 & & 65.0 & 28.4850 & & 600.0  & 127.4385 \\
16.0 & 2.9254 & & 70.0 & 31.0034 & & 650.0  & 128.7726 \\
17.0 & 3.4546 & & 75.0 & 33.5175 & & 700.0  & 129.8560 \\
18.0 & 4.0009 & & 80.0 & 36.0228 & & 800.0  & 131.4870 \\
19.0 & 4.5589 & & 90.0 & 40.9870 & & 900.0  & 132.6349 \\
20.0 & 5.1241 & & 100.0 & 45.8577 & & 1000.0  & 133.4713 \\
\hline\hline
\end{tabular}
\label{tab:HTO_C_v}
\end{table}


%

\end{document}